\documentclass{article}
\usepackage{spconf,amsmath,graphicx,hyperref}
\usepackage{tabularx} 
\usepackage{amsfonts}
\usepackage{graphicx}
\usepackage{makecell}
\usepackage{multirow} 
\usepackage{booktabs}
\usepackage{xcolor}
\usepackage[table]{xcolor}
\usepackage{threeparttable} 

\title{Joycent: Multi-Accent TTS via Disentangled Accent Modeling and Layer-Specific Conditioning}
\name{Xintong Wang, Junchuan Zhao, Ye Wang\thanks{This work is funded by a research grant
 MOE-MOESOL2021-0005 from the Ministry of Education in Singapore.}}
\address{School of Computing, National University of Singapore, Singapore}
\begin{document}
\ninept
\maketitle
\begin{abstract}
Accent text-to-speech (TTS) aims to synthesize speech with a target accent
while preserving speaker identity, but faces two key challenges:
disentangling accent from speaker characteristics and effectively
conditioning speech generation on the two disentangled factors.
In this paper, we propose Joycent, a diffusion-based accent TTS framework
that addresses both challenges.
Our key idea is to separate accent and speaker information in both
representation learning and TTS conditioning.
Joycent uses WhisAID, a Whisper-based accent encoder with gradient reversal
to learn speaker-disentangled accent representations, and introduces
layer-specific conditional layer normalization to inject accent and speaker
information at different stages of the text encoder.
We evaluate Joycent on the Mandarin Regional Accent Corpus (MRAC) with seen and unseen speakers, including a challenging cross-accent setting where the speaker and accent prompts come from different accents.  
Experimental results show that Joycent improves accent similarity over
existing methods while maintaining strong speaker similarity, with
consistent gains under the challenging cross-accent setting.
Subjective evaluation further confirms improved naturalness, accent
similarity, and speaker preservation. The audio samples are available at https://oshindow.github.io/joycent/.
\end{abstract}
\section{Introduction}
\label{sec:intro}
Text-to-speech (TTS) systems~\cite{du2024cosyvoice2scalablestreaming,10842513,DBLP:conf/icml/PopovVGSK21,DBLP:conf/icml/JuWS0XYLLST000024,deng2025indextts,zhou2025indextts2,wang2024maskgct} have made substantial progress in generating natural and expressive speech from text. Accent TTS~\cite{DBLP:conf/icassp/InoueWW0B025,ma2023accentvitsaccenttransferendtoendtts,zhang22_interspeech,halychanskyi2026fewshotaccentsynthesisasr,Xinyuan2025ScalableCA} further extends this capability to diverse regional and non-native accents. Recent systems increasingly use reference/prompt speech to specify the desired
accent and speaker characteristics, enabling flexible accent controls.
Speech reference/prompt-based accent TTS~\cite{xinyuan2025scalablecontrollableaccentedtts,DBLP:conf/icassp/ZhongRSS25,DBLP:conf/interspeech/BadlaniVSSGC23,deja2023diffusion} typically uses two speech prompts to independently
specify the target accent and speaker timbre: an accent prompt provides the
desired accent, while a speaker prompt provides the target timbre.
A particularly challenging case is cross-accent synthesis, where these prompts come from different speakers with different accents.
In this setting, the model must preserve the timbre of a typically unseen
target speaker while preventing the inherent accent of the speaker prompt
from interfering with the accent specified by the accent prompt.

The first challenge in cross-accent synthesis lies in disentangling accent from speaker identity.
In multi-accent corpora, each accent is typically represented by multiple speakers, whereas an individual speaker usually exhibits only one inherent accent~\cite{ma2023accentvitsaccenttransferendtoendtts,DBLP:journals/corr/abs-2410-13342}. Accent and speaker identity are therefore strongly correlated, allowing an accent encoder to retain speaker-dependent information that is predictive of the accent label. Related speech reference-based standard TTS face similar factor-entanglement issues and employ various forms of representation separation~\cite{DBLP:conf/interspeech/ZhaoWW25,zhao2026comelsinger}. In accent TTS, existing approaches either implicitly separate accent and speaker information through dedicated representations~\cite{DBLP:journals/corr/abs-2410-13342} or explicitly suppress speaker information from the accent representation using adversarial training~\cite{DBLP:conf/icassp/ZhongRSS25,DBLP:conf/interspeech/BadlaniVSSGC23}. We follow the latter direction and learn accent representations from pre-trained speech features, while explicitly reducing residual speaker information.

Further, even with a disentangled accent representation, it is imperative for an accent TTS model to know how to condition the accent information. Speech reference/prompt-based conditioning has been explored for speaking style~\cite{DBLP:conf/icml/WangSZRBSXJRS18,zhao2025spsinger,zhang2024stylesinger}, emotion~\cite{wang2023neural,liang2026ted}, and prosody~\cite{DBLP:conf/interspeech/ZhaoWW25}. In accent TTS, accent information is commonly introduced at the text-encoder input or on the decoder side. However, accent and speaker representations encode distinct speech factors and need not influence all stages of the linguistic representation in the same manner. We therefore investigate layer-specific conditioning, allowing accent and speaker information to modulate different stages of the text encoder for stronger accent control while preserving speaker identity.

In this work, we propose \textbf{Joycent}, a diffusion-based framework for cross-accent TTS that aims to address these challenges. Joycent learns a transferable accent representation with \textbf{WhisAID}, a Whisper-based~\cite{DBLP:conf/icml/RadfordKXBMS23} accent identification model with speaker suppression using gradient reversal~\cite{DBLP:conf/icml/GaninL15}, and combines it with an independently extracted speaker representation by FACodec~\cite{DBLP:conf/icml/JuWS0XYLLST000024} through layer-specific conditional layer normalization (CLN)~\cite{DBLP:conf/iclr/Chen0LLQZL21}. Joycent is trained and evaluated on 138 hours of Mandarin Regional Accent Corpus (MRAC) spanning 9 regional accents under seen-speaker, unseen-speaker \textbf{inherent-accent}, and unseen-speaker \textbf{cross-accent} settings. In the inherent-accent setting, the accent and speaker prompts come from different speakers but share the same inherent accent. In the cross-accent setting, they come from different speakers with different inherent accents. The latter directly evaluates whether the target accent can be transferred across speakers while preserving the identity specified by the speaker prompt. Objective and subjective evaluations demonstrate the effectiveness of Joycent.

Our main contributions are as follows:
\begin{itemize}
\item We study \textbf{accent--speaker disentanglement} in prompt-based accent TTS and introduce WhisAID, which learns extract accent representations while suppressing speaker
information.

\item We propose a \textbf{layer-specific conditioning strategy} that incorporates accent and speaker information at different text-encoder stages, improving accent control while preserving target-speaker characteristics.

\item We construct the 138 hours Mandarin Regional Accent Corpus (MRAC) dataset \textbf{spanning 9 regional mandarin accents} for training and evaluation. We further conduct a systematic evaluation under seen-speaker, unseen-speaker inherent-accent, and \textbf{unseen-speaker cross-accent settings,} with the latter explicitly evaluating accent transfer between speakers with different accents.

\end{itemize}

\section{Related Work}

\subsection{Accented Text-to-Speech}

Accented TTS aims to control accent characteristics while preserving linguistic
content and, in multi-speaker settings, the identity of a target speaker.
Existing methods differ primarily in how accent information is represented and
introduced into synthesis. DDGM-Acc~\cite{deja2023diffusion} formulate
accent modelling within a diffusion-based TTS framework and further perform
accent conversion by modifying accent-related features during the diffusion
process. DART~\cite{DBLP:journals/corr/abs-2410-13342} explicitly factorizes
accent and speaker representations using a multi-level variational autoencoder
with vector quantization, enabling different accent--speaker combinations at
inference time. AccentBox~\cite{DBLP:conf/icassp/ZhongRSS25} adopts a two-stage
pipeline: it first trains an accent identification model to extract
speaker-agnostic accent embeddings and then conditions a zero-shot TTS system
on these pretrained accent representations. These approaches demonstrate the
importance of separating accent from speaker identity when accent and speaker
conditions are recombined across utterances or speakers.

Other approaches model accent variation at a finer linguistic level.
Accent-VITS~\cite{ma2023accentvitsaccenttransferendtoendtts} decomposes the
text-to-wave mapping into text-to-accent and accent-to-wave stages and uses a
hierarchical CVAE to model accent pronunciation and acoustic characteristics.
Multi-scale accent modelling~\cite{zhou2025multiscaleaccentmodelingdisentangling}
captures both utterance-level accent characteristics and phoneme-level accent
variation while introducing speaker disentanglement for independent accent
control. Related methods explicitly predict or transform accent-dependent
phoneme representations before acoustic generation~\cite{
DBLP:conf/icassp/InoueWW0B025,
DBLP:journals/taslp/ZhouZZWL24}.
More recently, instruction-following TTS systems allow accent or dialect
characteristics to be specified through natural-language
instructions~\cite{du2025cosyvoice,chen2026flexivoice}.
In contrast, Joycent focuses on reference-based cross-accent synthesis, where
accent and speaker identity are provided by separate speech prompts, without
requiring explicit accented-phone prediction or textual accent descriptions.

\subsection{Speech Attribute Disentanglement}

Speech representations naturally mix multiple factors, including linguistic
content, speaker timbre, prosody, pitch, rhythm, style, emotion, and accent.
A major line of work therefore explicitly factorizes speech into
attribute-specific representations. SpeechSplit~\cite{DBLP:conf/icml/QianZCHC20}
uses multiple information bottlenecks to decompose speech into linguistic
content, timbre, pitch, and rhythm. NaturalSpeech~3~\cite{DBLP:conf/icml/JuWS0XYLLST000024}
extends this idea to neural speech codecs, where FACodec factorizes speech into
content, prosody, timbre, and acoustic-detail subspaces using factorized vector
quantization. Related work has also isolated prosodic information from other
speech factors for controllable voice conversion~\cite{DBLP:conf/interspeech/ZhaoWW25}
and separated style-related information from speaker characteristics in
expressive speech synthesis~\cite{DBLP:conf/icassp/ChoiBLMLC024}.
Together, these methods seek to assign different controllable attributes to
distinct latent spaces and thereby reduce interference between them.

A complementary strategy is to learn a representation for one target attribute
while explicitly suppressing nuisance information, rather than factorizing all
speech factors. AccentBox~\cite{DBLP:conf/icassp/ZhongRSS25}, for example,
learns speaker-agnostic accent representations using an information bottleneck
together with adversarial speaker suppression. In cross-speaker emotional TTS,
DiEmo-TTS~\cite{DBLP:conf/interspeech/ChoOKL25} instead uses self-supervised
distillation to obtain emotion representations with reduced speaker leakage.
Such suppression is particularly relevant to accented TTS when speaker identity
and accent are correlated in the training data, since a reference-derived
accent representation may otherwise retain speaker-specific cues. Joycent
follows this formulation: WhisAID is trained to preserve
accent-discriminative information while suppressing speaker-dependent
information, and the resulting accent representation is combined with an
independently extracted speaker representation for cross-accent synthesis.

\section{Method}
\label{sec:method}


\begin{figure*}[!t]
    \centering
    \includegraphics[width=\textwidth]{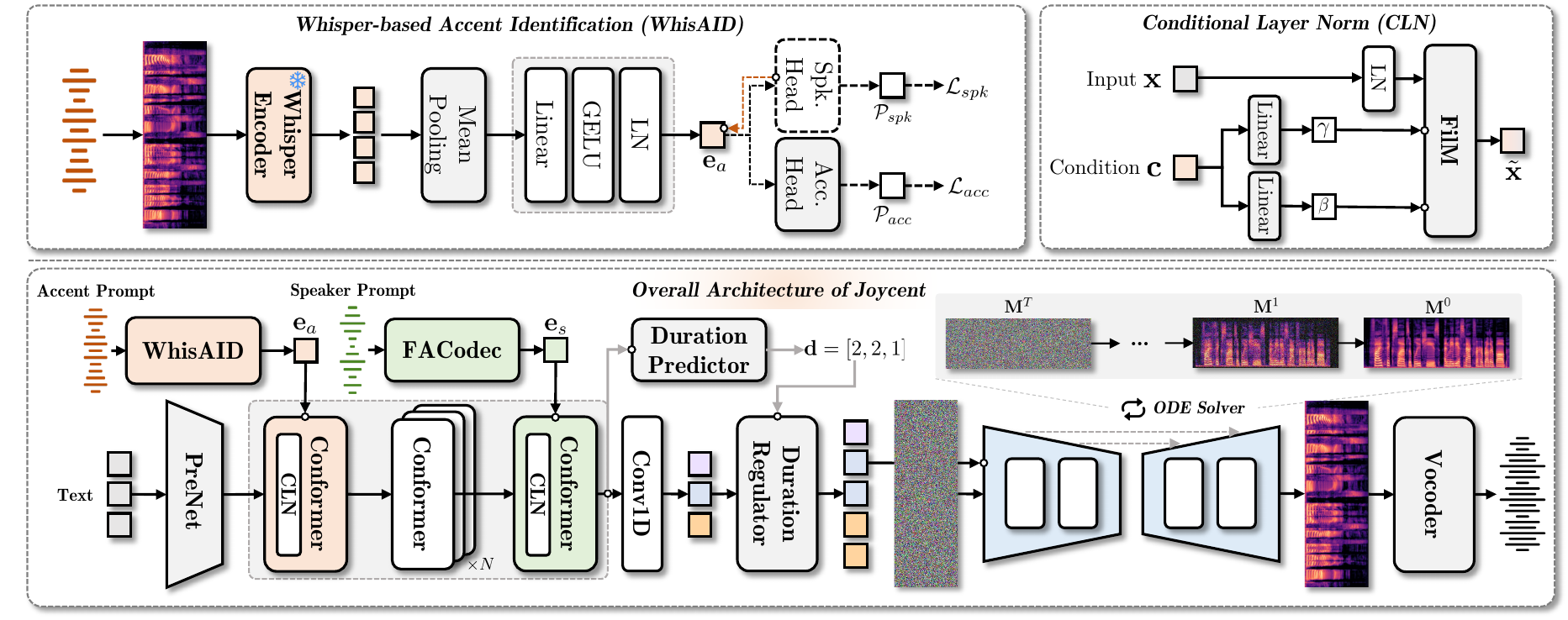}
    \caption{Overall architecture of \textbf{Joycent}. \textbf{WhisAID} learns speaker-disentangled accent representations, while CLN enables layer-specific accent and speaker conditioning. Joycent integrates these representations with the text encoder and generates speech through duration modeling and diffusion-based acoustic synthesis. The orange dashed lines indicate gradient reversal.}
    \label{fig:framework}
\end{figure*}

\subsection{Overall Architecture}
\label{sec:overall_architecture}

Given a text sequence $\mathbf{x}_{1:L}$, an accent prompt $\mathbf{w}_{a}$, and a speaker prompt $\mathbf{w}_{s}$, Joycent synthesizes speech that follows the accent specified by $\mathbf{w}_{a}$ while preserving the speaker identity specified by $\mathbf{w}_{s}$. As illustrated in Fig.~\ref{fig:framework}, the accent prompt is processed by our proposed WhisAID to obtain an accent representation $\mathbf{e}_{a}$. Independently, the speaker prompt is encoded using the timbre extractor of the pre-trained FACodec\footnote{FACodec: https://huggingface.co/amphion/naturalspeech3\_facodec} from NaturalSpeech 3~\cite{DBLP:conf/icml/JuWS0XYLLST000024} to obtain the speaker embedding $\mathbf{e}_{s}$. This separation allows the accent and speaker prompts to come from different speakers, including speakers with different inherent accents in the cross-accent setting. For linguistic encoding, $\mathbf{x}_{1:L}$ is first mapped to hidden representations by a PreNet and then processed by a Conformer-based encoder~\cite{gulati20_interspeech} with layer-specific conditioning. The pre-net consists of 3 Conv1D layers, each followed by layer normalization, ReLU activation, and dropout, followed by a linear projection and a residual connection. The first Conformer block is conditioned on $\mathbf{e}_{a}$ through conditional layer normalization (CLN), followed by $K$ standard Conformer blocks, while the final block is conditioned on $\mathbf{e}_{s}$ through another CLN module. This ordering introduces accent information early into the linguistic representation and speaker characteristics at a later stage, facilitating accent control while preserving speaker identity. A Conv1D projection produces the encoder representation $\mathbf{h}_{1:L}$, from which the duration predictor estimates $\mathbf{d}_{1:L}=[d_1,\ldots,d_L]$.

The duration regulator expands $\mathbf{h}_{1:L}$ according to $\mathbf{d}_{1:L}$ to obtain the frame-level acoustic condition $\mathbf{c}_{1:N}$. The expanded representation $\mathbf{c}_{1:N}$ is then provided to a score-based acoustic decoder following Grad-TTS~\cite{DBLP:conf/icml/PopovVGSK21}. Starting from Gaussian noise, the decoder iteratively generates the mel-spectrogram $\hat{\mathbf{M}}$ through the reverse diffusion process. Finally, a Parallel-WaveGAN vocoder~\footnote{Parallel WaveGAN: https://github.com/kan-bayashi/ParallelWaveGAN}~\cite{DBLP:conf/icassp/YamamotoSK20} converts $\hat{\mathbf{M}}$ into the synthesized waveform $\hat{\mathbf{w}}$. The following sections present the details of WhisAID, the layer-specific conditioning strategy for the Conformer-based encoder, and the training pipeline of Joycent.

\subsection{Whisper-based Accent Identification (WhisAID)}
\label{sec:whisaid}

Prompt-based accent synthesis requires an accent representation that captures accent-dependent pronunciation characteristics while minimizing information tied to the prompt speaker. Inspired by AccentBox~\cite{DBLP:conf/icassp/ZhongRSS25}, we introduce \textit{WhisAID}, a Whisper-based accent identification model with explicit speaker suppression. Whisper~\cite{DBLP:conf/icml/RadfordKXBMS23} provides rich phonetic and linguistic representations from large-scale speech recognition pre-training, making it well suited for modeling accent-dependent pronunciation patterns. Given an accent prompt $\mathbf{w}_{a}$, the waveform is processed by the standard Whisper feature extractor and a frozen pre-trained Whisper encoder to obtain the frame-level representation $\mathbf{h}_{a} \in \mathbb{R}^{N' \times D}$, where $N'$ is the number of encoded frames and $D$ is the feature dimension. We apply temporal mean pooling followed by a bottleneck projection consisting of a linear layer, GELU activation~\cite{hendrycks2023gaussianerrorlinearunits}, and layer normalization, producing the accent embedding $\mathbf{e}_{a} \in \mathbb{R}^{1 \times D'}$. The bottleneck compresses the Whisper representation into a compact space and limits the retention of nuisance information. A linear accent classifier is then applied to $\mathbf{e}_{a}$ and optimized with the cross-entropy loss $\mathcal{L}_{\mathrm{acc}}$.

To explicitly reduce speaker information in $\mathbf{e}_{a}$, we attach an auxiliary linear speaker classifier through a gradient reversal layer (GRL)~\cite{DBLP:conf/icml/GaninL15}. The GRL reverses the speaker-classification gradient propagated to the accent representation, encouraging the bottleneck to preserve accent-discriminative information while suppressing speaker-dependent cues. The speaker branch is optimized using the cross-entropy loss $\mathcal{L}_{\mathrm{spk}}$, and the overall WhisAID objective is: 
\begin{equation}
\mathcal{L}_{\mathrm{WhisAID}}
=
\mathcal{L}_{\rm{acc}}
+
\lambda \mathcal{L}_{\rm{spk}},
\end{equation}
where $\lambda$ controls the strength of adversarial speaker suppression. During training, the Whisper encoder remains frozen, while the bottleneck and both classification heads are jointly optimized. At inference time, the classification heads and GRL branch are removed, and the Whisper encoder together with the bottleneck projection is used to extract $\mathbf{e}_{a}$ for conditioning Joycent.

\subsection{Layer-specific Accent and Speaker Conditioning}

\label{sec:layer_conditioning}

Existing accent TTS systems commonly incorporate accent information by concatenating or adding accent embeddings to speaker representations. In contrast, Joycent introduces accent and speaker information at different stages of the text encoder through conditional layer normalization (CLN). Specifically, the layer normalization following the first Conformer block is replaced with an accent-conditioned CLN, while the layer normalization after the final Conformer block is replaced with a speaker-conditioned CLN. We introduce accent conditioning early in the encoder since accent variations are closely related to phone realization~\cite{ezzerg2023remap,yan2004analysis}, allowing accent information to modulate token-level linguistic representations before higher-level acoustic characteristics are formed. Speaker information is introduced at the final encoder stage to preserve the target speaker characteristics after accent-conditioned linguistic modeling.

For each CLN, the conditioning embedding is linearly projected into feature-wise affine parameters $\gamma(\cdot)$ and $\beta(\cdot)$, which modulate the normalized hidden representation. Let \(\mathbf{h}_a\) and \(\mathbf{h}_s\) denote the representations obtained by broadcasting the utterance-level accent and speaker embeddings \(\mathbf{e}_a\) and \(\mathbf{e}_s\) to match the text token sequence length $L$, respectively. The corresponding conditioning operations are implemented using
Feature-wise Linear Modulation (FiLM)~\cite{perez2018film}
\begin{equation}
\begin{aligned}
\tilde{\mathbf{h}}_a &=
\gamma(\mathbf{e}_a)
\odot
\frac{\mathbf{h}_a-\mu(\mathbf{h}_a)}
{\sigma(\mathbf{h}_a)}
+
\beta(\mathbf{e}_a), \\
\tilde{\mathbf{h}}_s &=
\gamma(\mathbf{e}_s)
\odot
\frac{\mathbf{h}_s-\mu(\mathbf{h}_s)}
{\sigma(\mathbf{h}_s)}
+
\beta(\mathbf{e}_s),
\end{aligned}
\end{equation}
where $\mu(\cdot)$ and $\sigma(\cdot)$ denote the mean and standard deviation, respectively, and $\odot$ denotes element-wise multiplication. This layer-specific design allows accent and speaker information to influence different stages of linguistic encoding rather than being fused through a shared conditioning pathway. As shown in our ablation experiments, conditioning accent information at an early Conformer block yields stronger accentedness than conditioning the frame-level diffusion decoder, supporting the importance of introducing accent information during linguistic encoding.
 
\begin{table*}[!t]
\centering
\caption{
Statistics of the Mandarin multi-accent dataset.
Seen and Unseen denote the seen- and unseen-speaker test sets, respectively.
Speakers in the Seen split overlap with those in the training set.
}
\label{tab:dataset_statistics}

\small
\setlength{\tabcolsep}{3.8pt}
\renewcommand{\arraystretch}{1.12}

\begin{tabular}{@{}l
rrr
rrrr
rrr@{}}
\toprule
&
\multicolumn{3}{c}{\textbf{Utterances}}
&
\multicolumn{4}{c}{\textbf{Duration (h)}}
&
\multicolumn{3}{c}{\textbf{Speakers}}
\\
\cmidrule(lr){2-4}
\cmidrule(lr){5-8}
\cmidrule(l){9-11}

\textbf{Accent}
& \textbf{Train}
& \textbf{Seen Test}
& \textbf{Unseen Test}
& \textbf{Train}
& \textbf{Seen Test}
& \textbf{Unseen Test}
& \textbf{Total}
& \textbf{Train}
& \textbf{Seen Test}
& \textbf{Unseen Test}
\\
\midrule

North
& 43,467 & 2,000 & 2,000
& 44.307 & 1.672 & 1.931 & 47.910
& 126 & 106 & 39 \\

Sichuan
& 19,066 & 2,377 & 2,377
& 16.356 & 2.081 & 1.801 & 20.238
& 49 & 49 & 2 \\

Guangdong
& 19,029 & 2,371 & 2,371
& 15.467 & 1.930 & 1.409 & 18.806
& 51 & 51 & 2 \\

South
& 17,651 & 1,000 & 1,000
& 16.667 & 0.919 & 0.840 & 18.426
& 44 & 42 & 7 \\

Henan
& 14,236 & 1,769 & 1,769
& 10.269 & 1.271 & 1.314 & 12.854
& 20 & 20 & 2 \\

Shanghai
& 7,603 & 648 & 648
& 6.923 & 0.635 & 0.541 & 8.098
& 28 & 28 & 2 \\

Wuhan
& 4,073 & 461 & 461
& 4.178 & 0.483 & 0.383 & 5.044
& 9 & 9 & 1 \\

Tianjin
& 6,552 & 859 & 859
& 2.792 & 0.354 & 0.514 & 3.659
& 6 & 6 & 1 \\

Singapore
& 3,729 & 1,109 & 1,109
& 2.218 & 0.660 & 0.522 & 3.400
& 3 & 3 & 1 \\

\midrule
\textbf{Total}
& \textbf{135,406}
& \textbf{12,594}
& \textbf{12,594}
& \textbf{119.178}
& \textbf{10.004}
& \textbf{9.254}
& \textbf{138.435}
& \textbf{336}
& \textbf{314}
& \textbf{57}
\\
\bottomrule
\end{tabular}
\end{table*}

\subsection{Training Strategy}
\label{sec:training_strategy}

Joycent is trained in two stages. In the first stage, WhisAID is trained independently to learn accent representations that preserve accent-discriminative information while suppressing speaker-dependent cues. The pre-trained Whisper encoder remains frozen, while the bottleneck projection and the accent and speaker classification heads are jointly optimized, following the objectives described in Section~\ref{sec:whisaid}. After convergence, the classification heads are
discarded, while the WhisAID encoder and bottleneck projection are frozen and used to extract the accent embedding $\mathbf{e}_{a}$ from
the accent prompt, while the frozen pre-trained
FACodec timbre extractor extracts the speaker embedding $\mathbf{e}_{s}$.

In the second stage, the PreNet, Conformer-based text encoder, duration
predictor, and Conv1D layer producing the hidden representation
$\mathbf{h}_{1:L}$ are jointly optimized from the text input, with
$\mathbf{e}_{a}$ and $\mathbf{e}_{s}$ providing the accent and speaker
conditions, respectively. Following Grad-TTS~\cite{DBLP:conf/icml/PopovVGSK21}, we employ monotonic alignment search (MAS)~\cite{DBLP:conf/nips/KimKKY20} to align the token-level encoder representations $\mathbf{h}_{1:L}$ with the ground-truth mel-spectrogram $\mathbf{M}_{1:N}$. Let $\mathbf{A} \in \{0,1\}^{L\times N}$ denote a monotonic alignment matrix, where $\mathbf{A}_{l,n}=1$ indicates that the $n$-th acoustic frame is aligned to the $l$-th text token. MAS determines the optimal alignment as: 
\begin{equation}
\mathbf{A}^{*}
=
\arg\max_{\mathbf{A}\in\mathcal{A}}
\sum_{l=1}^{L}
\sum_{n=1}^{N}
A_{l,n}
\log
\mathcal{N}
\left(
\mathbf{M}_{n};
\mathbf{h}_{l},
\mathbf{I}
\right),
\end{equation}
where $\mathcal{A}$ denotes the set of valid monotonic alignments. Based on $\mathbf{A}^{*}$, the token-level representations are expanded to the frame-level acoustic condition $\mathbf{c}_{1:N}$, and the corresponding Gaussian prior is optimized by: 
\begin{equation}
\mathcal{L}_{\mathrm{MAS}}
=
-\frac{1}{N}
\sum_{n=1}^{N}
\log
\mathcal{N}
\left(
\mathbf{M}_{n};
\mathbf{c}_{n},
\mathbf{I}
\right).
\end{equation}
The same alignment provides supervision for the duration predictor. The target duration of the $l$-th token is obtained as
$d_l^{*}
=
\sum_{n=1}^{N} A^{*}_{l,n}$, 
and the corresponding log-duration prediction $\hat{d}_l$ is optimized with: 
\begin{equation}
\mathcal{L}_{\mathrm{dur}}
=
\frac{1}{L}
\sum\limits_{l=1}^{L}
\left(
\hat{d}_l
-
\log(d_l^{*}+1)
\right)^2.
\end{equation}
Given the aligned acoustic condition $\mathbf{c}_{1:N}$, the score-based decoder is trained using the diffusion objective $\mathcal{L}_{\mathrm{diff}}$. The overall Stage two training objective is therefore: 
\begin{equation}
\mathcal{L}_{\mathrm{TTS}}
=
\mathcal{L}_{\mathrm{MAS}}
+
\mathcal{L}_{\mathrm{dur}}
+
\mathcal{L}_{\mathrm{diff}}.
\end{equation}


\section{Experimental Setup}
\label{sec:exp}
\subsection{Datasets}
\label{sec:dataset}

We construct the \textbf{Mandarin Regional Accent Corpus (MRAC) }from 7 regional Mandarin accent categories,
including Guangdong, Shanghai, Sichuan, Tianjin, Henan, Wuhan, and
Singapore, collected from the MagicHub platform\footnote{MagicHub:
https://magichub.com/datasets/}, together with 2 accents from the
AISHELL-3~\cite{DBLP:conf/interspeech/ShiBXZL21} corpus. For the 7 regional accents, we construct seen-speaker and unseen-speaker
test sets to evaluate both inherent-accent and cross-accent speech
synthesis. Specifically, 2 speakers are randomly selected for
unseen-speaker evaluation for each of Guangdong, Shanghai, Sichuan, and
Henan. For Tianjin, Wuhan, and Singapore, 1 speaker is randomly selected for unseen-speaker
evaluation, each of which contains no more
than 10 speakers. All utterances from these reserved speakers form the
unseen-speaker test set. From the remaining speakers, we randomly sample
the same number of utterances for each accent to construct the
seen-speaker test set, with all remaining utterances used for training. For AISHELL-3, we exclude speakers labeled as ``others'' and group the
remaining speakers into the North and South accent categories according
to the corpus annotations. All evaluation utterances are selected from
the original AISHELL-3 test partition. For the North category, 2,000
utterances from 39 speakers are randomly selected for unseen-speaker
evaluation. For the South category, due to the limited amount of
available data, 1,000 utterances from 7 speakers are randomly selected
for unseen-speaker evaluation. We then randomly sample the same number
of utterances from speakers appearing in the training set to construct
the corresponding seen-speaker test sets. The resulting dataset contains 9 accent categories and 138 hours of
speech in total. Detailed statistics for each accent category are reported
in Table~\ref{tab:dataset_statistics}.

\subsection{Experimental Settings}

\textbf{WhisAID.} We implement WhisAID using the pre-trained Whisper-medium encoder~\footnote{Whisper-Medium: https://huggingface.co/openai/whisper-medium} with 80-dimensional log Mel-spectrograms as input. All audio segments are zero-padded to 30 seconds. The feature dimension \(D\) is determined at 1024 by the Whisper, while we set the projected dimension \(D'=256\). The coefficient $\lambda$ for $\mathcal{L}_{\rm spk}$ is empirically set to 0.05 based on the results in Table~\ref{tab:whisaid}. WhisAID is trained for 10 epochs with a batch size of 32 using AdamW~\cite{DBLP:conf/iclr/LoshchilovH19}, a weight decay of 0.01, and a cosine learning rate schedule with a peak learning rate of $1\times10^{-5}$ and 1,000 warm-up steps.


\textbf{Joycent.}
The input text is converted to tone-marked Pinyin using
G2PW\footnote{G2PW: https://github.com/GitYCC/g2pW} for polyphone
disambiguation. Each syllable is then decomposed into an initial and a
tone-marked final, while zero-initial syllables are kept as single tokens.
Silence tokens are added at both ends, with blank tokens inserted between
adjacent non-silence tokens for monotonic alignment.
We use $K=4$ Conformer blocks with a dropout rate of 0.1 and follow the
Grad-TTS~\cite{DBLP:conf/icml/PopovVGSK21} configuration for the score-based
decoder. Joycent is trained from scratch for 400k steps with a batch size of
16 using Adam~\cite{DBLP:journals/corr/KingmaB14} and a constant learning rate
of $1\times10^{-4}$, and uses 10 reverse diffusion steps at inference.
All audio is resampled to 16~kHz and represented as 80-dimensional log-Mel
spectrograms using a 1,024-point FFT, a hop size of 256, and a Mel range of
80--7,600~Hz, following the Parallel WaveGAN acoustic configuration.
A Parallel WaveGAN vocoder is trained for 400k steps with a batch size of 6
on CSMSC\footnote{CSMSC: https://www.data-baker.com/open\_source.html} and
the datasets listed in Table~\ref{tab:dataset_statistics}.



\subsection{Evaluation Metrics}

\textbf{Objective evaluation.}
We evaluate the WhisAID and Joycent separately. For WhisAID, following GenAID~\cite{DBLP:conf/icassp/ZhongRSS25}, we
report \textbf{accuracy} and \textbf{F1} for seen speakers, and additionally \textbf{precision} and
\textbf{recall} for unseen speakers.
Speaker information retained in the accent embeddings is measured using the
\textbf{Silhouette Coefficient for Speaker Clusters} (\textbf{SCSC}), where a lower value
indicates better speaker disentanglement, together with a linear speaker probe
trained on frozen accent embeddings, for which precision, recall, F1, and
accuracy are reported. For Joycent, we evaluate synthesized speech under seen-speaker inherent-accent (Seen-Inherent),
unseen-speaker inherent-accent (Unseen-Inherent), and unseen-speaker cross-accent (Unseen-Cross) settings. Speaker similarity \textbf{(Spk.-CS)} is measured by the cosine similarity between
WavLM-Large speaker embeddings\footnote{WavLM-Large: https://huggingface.co/microsoft/wavlm-large},
while intelligibility is evaluated using \textbf{Character Error Rate} (\textbf{CER}) obtained with
Qwen3-ASR-1.7B\footnote{Qwen3-ASR-1.7B: https://huggingface.co/Qwen/Qwen3-ASR-1.7B}.
Following AccentBox~\cite{DBLP:conf/icassp/ZhongRSS25}, accent similarity is
measured by the cosine similarity between synthesized and prompt accent
embeddings extracted using both WhisAID (\textbf{W-CS}) and WhisAID$_{\rm adv.}$ (\textbf{WU-CS}), reducing
evaluation bias toward a single accent representation.

\textbf{Subjective evaluation.}
We conduct listening tests on ground-truth and synthesized speech. 15 native Chinese listeners familiar with the evaluated accents rate speech naturalness on a 5-point \textbf{MOS} scale. Accent and speaker similarity are evaluated separately using 4-point similarity MOS (\textbf{SMOS-A} and \textbf{SMOS-S}), where listeners rate the similarity of each synthesized sample to the target-accent and target-speaker prompt speech, respectively.

\section{Results}
\label{sec:results}
\subsection{WhisAID}

\begin{table*}[!t]
\centering
\caption{Objective evaluation and ablation results of WhisAID.
Results are reported with ablations on the
disentanglement method, speaker classification loss weight $\lambda$, and Whisper encoder scale.
Prec., Rec., and Acc. denote macro-averaged precision, recall, and accuracy, respectively.
Best and second-best results are shown in \textbf{bold} and \underline{underlined}, respectively.}
\label{tab:whisaid}
\setlength{\tabcolsep}{4pt}
\renewcommand{\arraystretch}{1.05}
\begin{tabular}{llcc c cc cccc c}
\toprule
\multirow{2}{*}{System}
& \multirow{2}{*}{Encoder}
& \multirow{2}{*}{Method}
& \multirow{2}{*}{$\lambda$}
& \multicolumn{2}{c}{Seen speakers $\uparrow$}
& \multicolumn{4}{c}{Unseen speakers $\uparrow$}
& \multirow{2}{*}{SCSC $\downarrow$} \\
\cmidrule(lr){5-6}
\cmidrule(lr){7-10}
& & & & F1 & Acc. & Prec. & Rec. & F1 & Acc. & \\
\midrule

GenAID
& XLSR$_{\rm large}$
& Adv. & --
& \textbf{0.94} & \textbf{0.94}
& 0.58 & 0.59 & 0.57 & 0.59
& 0.26 \\

\midrule

\multirow{3}{*}{WhisAID}
& Whisper$_{\rm medium}$
& GRL & 0.10
& 0.89 & 0.88
& \underline{0.73} & 0.65 & 0.66 & 0.67
& 0.15 \\
& \cellcolor{gray!15}\textbf{Whisper$_{\rm medium}$}
& \cellcolor{gray!15}GRL & \cellcolor{gray!15}0.05
& \cellcolor{gray!15}0.89 & \cellcolor{gray!15}0.88
& \cellcolor{gray!15}0.72 & \cellcolor{gray!15}\underline{0.65} & \cellcolor{gray!15}\underline{0.66} & \cellcolor{gray!15}\underline{0.67}
& \cellcolor{gray!15}\underline{0.12} \\
& Whisper$_{\rm medium}$
& GRL & 0.01
& 0.89 & \underline{0.89}
& 0.73 & 0.65 & \textbf{0.67} & \textbf{0.68}
& 0.17 \\

\midrule

WhisAID$_{\rm adv.}$
& Whisper$_{\rm medium}$
& Adv. & --
& \underline{0.90} & 0.89
& \textbf{0.74} & 0.65 & 0.66 & 0.68
& 0.22 \\

WhisAID$_{\mathrm{w/o\,GRL}}$
& Whisper$_{\rm medium}$
& None & --
& 0.90 & 0.89
& 0.74 & \textbf{0.66} & 0.67 & 0.68
& 0.28 \\

\midrule

WhisAID$_{\rm small}$
& Whisper$_{\rm small}$
& GRL & 0.05
& 0.88 & 0.87
& 0.66 & 0.59 & 0.60 & 0.60
& 0.13 \\

WhisAID$_{\rm large\text{-}v3\text{-}turbo}$
& Whisper$_{\rm large-v3-turbo}$
& GRL & 0.05
& 0.85 & 0.85
& 0.67 & 0.62 & 0.63 & 0.64
& \textbf{0.09} \\

\bottomrule
\end{tabular}
\end{table*}

We use \textbf{GenAID}~\cite{DBLP:conf/icassp/ZhongRSS25} as the baseline. GenAID is based on XLSR-Large\footnote{XLSR: https://huggingface.co/facebook/wav2vec2-large-xlsr-53} and employs adversarial speaker disentanglement that encourages the speaker classifier to produce a uniform distribution over speaker identities. \textbf{WhisAID$_{\mathrm{adv.}}$ }replaces GRL with the adversarial training strategy used in GenAID. \textbf{WhisAID$_{\mathrm{w/o\,GRL}}$} removes speaker classifier and GRL-based speaker disentanglement. \textbf{WhisAID$_{\rm small}$} and \textbf{WhisAID$_{\rm large\text{-}v3\text{-}turbo}$} replace the Whisper-Medium encoder with Whisper-Small and Whisper-Large-v3-Turbo, respectively. We additionally vary the $\lambda$ coefficient from 0.01 to 0.10 to examine its effect on accent identification and speaker disentanglement. All baselines and ablations are trained from scratch using the same training data as WhisAID.

\subsubsection{Objective Evaluation}
As shown in Table~\ref{tab:whisaid}, GenAID achieves the best performance on seen speakers, with both F1 and accuracy reaching 0.94, but its performance drops substantially for unseen speakers (0.57 F1 and 0.59 accuracy). In contrast, WhisAID with the Whisper Medium encoder achieves slightly lower performance on seen speakers but substantially better performance on unseen speakers, reaching 0.67 F1 and 0.68 accuracy ($\lambda=0.01$). The smaller seen--unseen performance gap suggests that WhisAID generalizes better to speakers not observed during training. We next examine the effect of speaker disentanglement. WhisAID$_{\mathrm{w/o\,GRL}}$, which uses Whisper Medium as the encoder with GRL disabled, yields the highest SCSC of 0.28, indicating stronger speaker-dependent clustering in the learned accent representations. WhisAID$_{\rm adv.}$ reduces the SCSC to 0.22, while GRL with $\lambda=0.05$ further reduces it to 0.12 while maintaining comparable accent identification performance. These results show that GRL effectively reduces speaker-dependent clustering in the accent embedding space. Among the evaluated weights for $\mathcal{L}_{\rm spk}$, $\lambda=0.05$ achieves the lowest SCSC, compared with 0.15 and 0.17 for $\lambda=0.10$ and $0.01$, respectively, suggesting a better balance between accent identification and speaker disentanglement. 

Finally, we compare different Whisper encoder scales using $\lambda=0.05$. Whisper Medium achieves the best accent identification performance, with an unseen-speaker F1 of 0.66 and accuracy of 0.67. Increasing encoder size does not consistently improve performance: Whisper Large-v3-Turbo yields a lower SCSC of 0.09, indicating stronger speaker disentanglement, but lower unseen-speaker F1 and accuracy of 0.63 and 0.64. Whisper Small performs worse on unseen speakers. Overall, Whisper Medium provides the best balance between accent discrimination and speaker disentanglement. \textbf{We therefore adopt Whisper$_{\rm medium}$ as the encoder backbone of WhisAID in the final Joycent model.}

\subsubsection{Speaker-information Probing}
We further train a linear classifier for 10 epochs on the frozen accent embeddings to predict speaker identities. As shown in Table~\ref{tab:speaker_probe}, speaker classification performance is clearly reduced when using either adversarial training or GRL compared with the variant without GRL, whereas the difference between the two methods is relatively small. This suggests that the two adversarial approaches retain a similar degree of speaker information that remains linearly recoverable to some extent. Interestingly, GRL achieves a substantially lower SCSC than conventional adversarial training (0.12 v.s. 0.22) as shown in Table~\ref{tab:whisaid}. One possible explanation is that GRL more strongly alters the geometry of the accent embedding space, making the embeddings less clustered by speaker.

\begin{table}[!t]
\centering
\caption{Speaker probing results on WhisAID accent embeddings. Prec., Rec., and Acc. denote macro-averaged precision, recall, and accuracy, respectively.   Lower values indicate less speaker-discriminative information.}
\label{tab:speaker_probe}

\resizebox{0.75\linewidth}{!}{%
\begin{tabular}{lcccc}
\toprule
Model
& Prec. $\downarrow$
& Rec. $\downarrow$
& F1 $\downarrow$
& Acc. $\downarrow$ \\
\midrule

WhisAID$_{\mathrm{w/o\,GRL}}$
& 0.46 & 0.55 & 0.46 & 0.66 \\

WhisAID$_{\rm adv.}$
& 0.44  & 0.50& 0.42 & 0.64 \\

\cellcolor{gray!15}\textbf{WhisAID}
& \cellcolor{gray!15}\textbf{0.43} &\cellcolor{gray!15}\textbf{ 0.50} & \cellcolor{gray!15}\textbf{0.42} & \cellcolor{gray!15}\textbf{0.63} \\

\bottomrule
\end{tabular}%
}
\end{table}

\subsubsection{Representation Visualization} 
\begin{figure}[!t]
    \centering
    \includegraphics[width=0.9\linewidth]{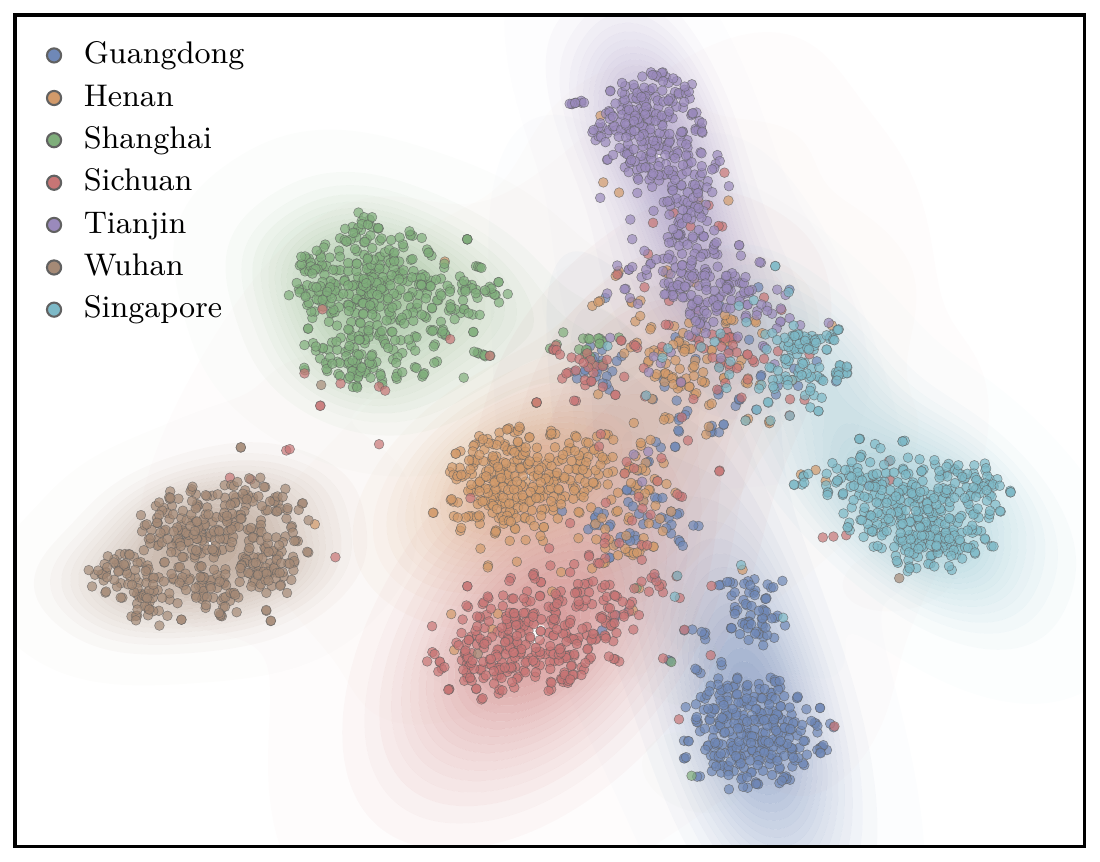}
    \caption{t-SNE visualization of the learned accent representations across different regional accents.}
    \label{fig:tsne_accent}
\end{figure}

\begin{figure}[!t]
    \centering
    \includegraphics[width=0.9\linewidth]{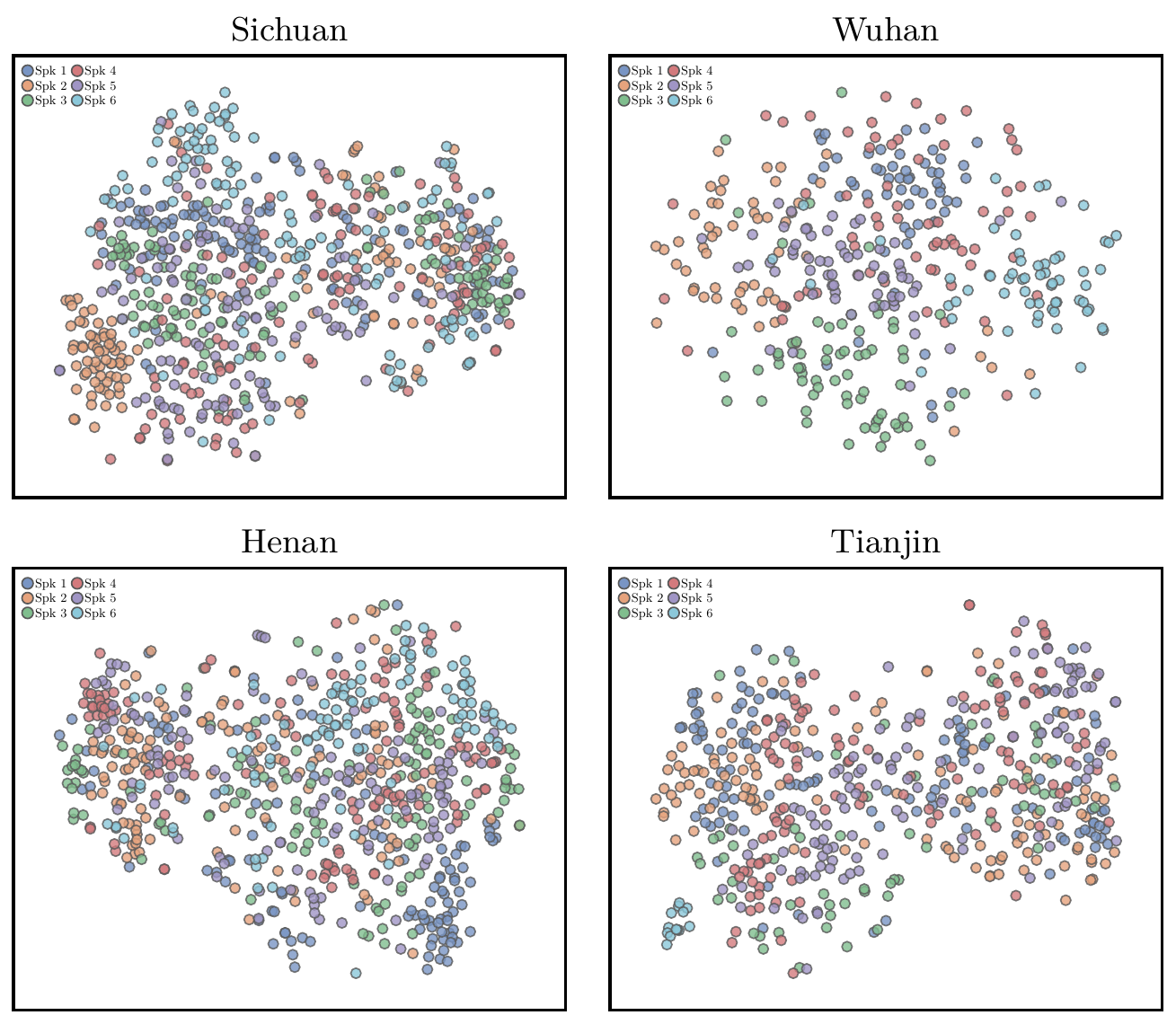}
    \caption{t-SNE visualization of the learned accent representations colored by speaker identity within different regional accents.}
    \label{fig:tsne_speaker}
\end{figure}

We further visualize the learned accent embeddings using t-SNE. Fig.~\ref{fig:tsne_accent} shows distinct clusters across regional accents, indicating that the learned representations capture accent-discriminative information. Fig.~\ref{fig:tsne_speaker} further visualizes the accent embeddings by speaker identity within each accent. The substantial overlap among speakers suggests that the learned accent representations do not form distinct speaker-dependent clusters.

\subsection{Accent TTS}

\begin{figure}[t]
    \centering
    \includegraphics[width=0.8\linewidth]{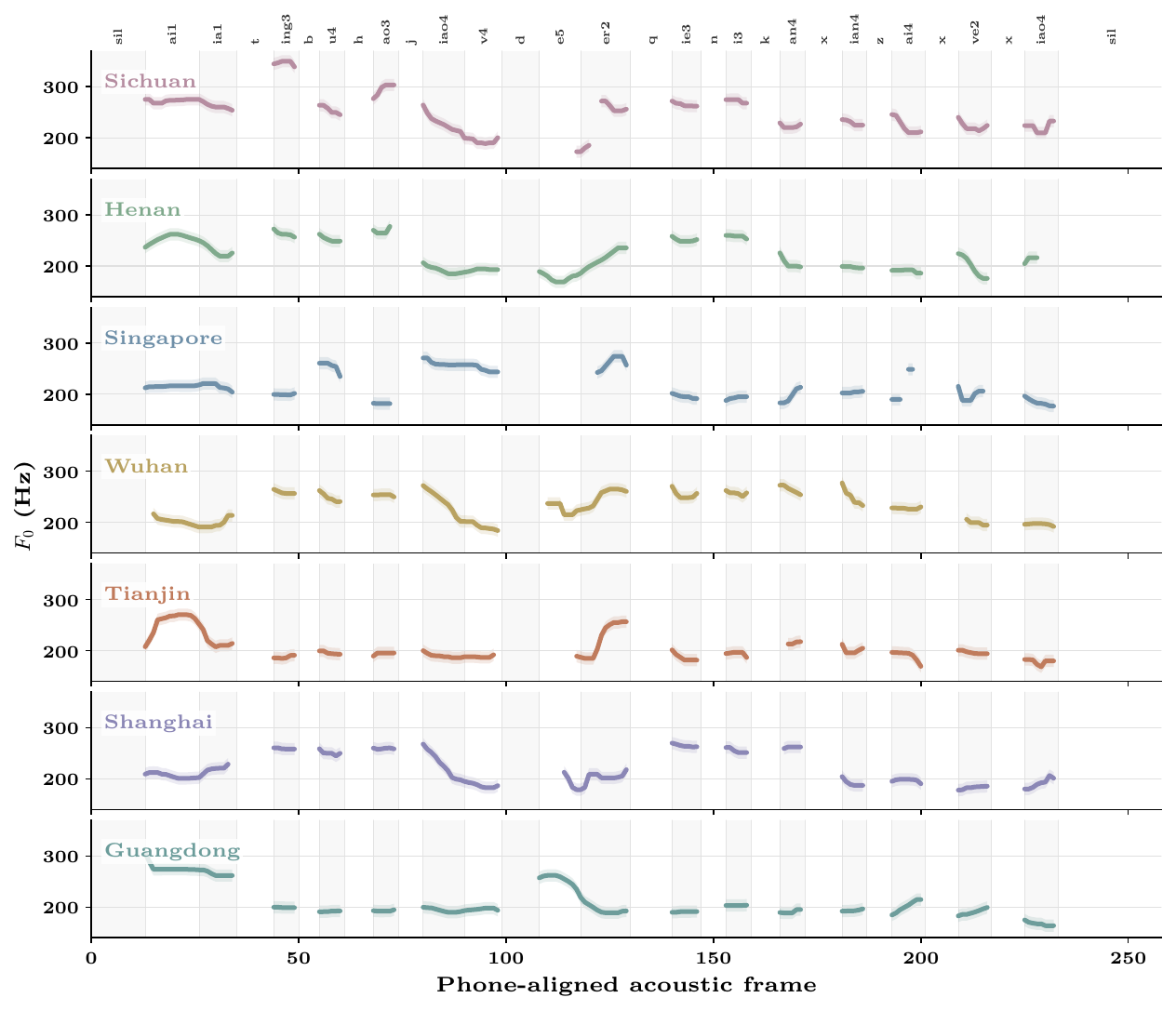}
    \caption{
Phone-aligned F0 contours are shown for speech synthesized using the same
text and speaker prompt under seven target accents. For visualization, the
F0 contour within each phone is time-aligned across the synthesized
samples.}
    \label{fig:f0_accent_comparison}
\end{figure}

We compare Joycent with three accent TTS baselines:
\textbf{DDGM-Acc}~\cite{deja2023diffusion},
\textbf{AccentBox}~\cite{DBLP:conf/icassp/ZhongRSS25}, and
\textbf{DART}~\cite{DBLP:journals/corr/abs-2410-13342}. All baselines are implemented following their original configurations and
trained on the same training data as Joycent from scratch. For ablation studies,
\textbf{Joycent$_{\rm small}$} and
\textbf{Joycent$_{\rm large\text{-}v3\text{-}turbo}$} are implemented using accent embeddings
extracted by WhisAID$_{\rm small}$ and
WhisAID$_{\rm large\text{-}v3\text{-}turbo}$, respectively.
\textbf{Joycent$_{\mathrm{w/o\,GRL}}$} uses accent embeddings extracted by
WhisAID$_{\mathrm{w/o\,GRL}}$.
\textbf{Joycent$_{\mathrm{w/o\,CLN}}$} replaces CLN-based accent and speaker conditioning with direct addition of the phone, accent, and speaker embeddings as input to the text encoder.
\textbf{Joycent$_{\rm acc\text{-}blk3}$} moves CLN-based accent conditioning
from the first to the third Conformer block while keeping speaker
conditioning unchanged.

\subsubsection{Quantitative Evaluation}

\begin{table*}[t]
\centering
\caption{Objective evaluation results of Joycent and baseline systems under three synthesis settings. \textbf{Bold}: best; \underline{underline}: second best among synthesized systems.}
\label{tab:tts_objective}
\setlength{\tabcolsep}{3pt}
\renewcommand{\arraystretch}{1.1}

\begin{tabular}{lcccccccccccc}
\toprule
\multirow{2}{*}{Model}
& \multicolumn{4}{c}{Seen-Inherent}
& \multicolumn{4}{c}{Unseen-Inherent}
& \multicolumn{4}{c}{Unseen-Cross} \\
\cmidrule(lr){2-5}
\cmidrule(lr){6-9}
\cmidrule(lr){10-13}
& W-CS $\uparrow$ & WU-CS $\uparrow$ & Spk.-CS $\uparrow$ & CER $\downarrow$
& W-CS $\uparrow$ & WU-CS $\uparrow$ & Spk.-CS $\uparrow$ & CER $\downarrow$
& W-CS $\uparrow$ & WU-CS $\uparrow$ & Spk.-CS $\uparrow$ & CER $\downarrow$ \\
\midrule

Ground Truth
& 0.530 & 0.563 & 0.574 & 0.085
& 0.577 & 0.611 & 0.584 & 0.123
& - & - & - & - \\

\midrule

DDGM-Acc~\cite{deja2023diffusion}
& 0.397 & 0.427 & \textbf{0.639} & 0.311
& \textbf{0.473} & \underline{0.510} & \underline{0.601} & 0.410
& \underline{0.382} & \underline{0.419} & \underline{0.567} & 0.285 \\

DART
& 0.321 & 0.357 & 0.555 & \textbf{0.153} & 0.350 & 0.403 & 0.557 & \textbf{0.165} & 0.344 & 0.387 & 0.524 & \textbf{0.150} \\

AccentBox
& \underline{0.404} & \underline{0.434} & 0.550 & 0.535
& 0.401 & 0.448 & 0.509 & 0.593
& 0.337 & 0.376 & 0.512 & 0.555 \\

\rowcolor{gray!15}
\textbf{Joycent (Ours)}
& \textbf{0.490} & \textbf{0.526} & \underline{0.626} & \underline{0.253}
& \underline{0.469} & \textbf{0.523} & \textbf{0.618} & \underline{0.302}
& \textbf{0.412} & \textbf{0.448} & \textbf{0.584} & \underline{0.258} \\

\bottomrule
\end{tabular}
\end{table*}

\begin{table*}[t]
\centering
\caption{Per-accent objective evaluation results of Joycent and baseline systems under unseen speaker settings. HN, SC, TJ, WH, and SG denote Henan, Sichuan,
Tianjin, Wuhan, and Singapore Mandarin, respectively.
}
\label{tab:tts_per_accent}
\setlength{\tabcolsep}{2.5pt}
\renewcommand{\arraystretch}{1.05}
\begin{tabular}{ll*{5}{cc}}
\toprule
\multirow{2}{*}{Target Speaker}
& \multirow{2}{*}{Model}
& \multicolumn{2}{c}{HN}
& \multicolumn{2}{c}{SC}
& \multicolumn{2}{c}{TJ}
& \multicolumn{2}{c}{WH}
& \multicolumn{2}{c}{SG} \\
\cmidrule(lr){3-4}
\cmidrule(lr){5-6}
\cmidrule(lr){7-8}
\cmidrule(lr){9-10}
\cmidrule(lr){11-12}
&
& W-CS & WU-CS
& W-CS & WU-CS
& W-CS & WU-CS
& W-CS & WU-CS
& W-CS & WU-CS \\
\midrule

\multirow{4}{*}{\shortstack{Unseen-Inherent}}
& DDGM-Acc~\cite{deja2023diffusion}
& \textbf{0.353} & \underline{0.364}
& \underline{0.508} & 0.525
& \textbf{0.608} & \textbf{0.663}
& \underline{0.314} & 0.356
& \textbf{0.580} & \textbf{0.643} \\

& DART
& 0.182 & 0.241
& 0.458 & 0.496
& 0.328 & 0.350
& 0.230 & 0.271
& 0.153 & 0.277 \\

& AccentBox
& 0.248 & 0.280
& \underline{0.508} & \underline{0.548}
& 0.508 & 0.529
& \underline{0.314} & \underline{0.371}
& 0.425 & 0.511 \\

& \cellcolor{gray!15}\textbf{Joycent (Ours)}
& \cellcolor{gray!15}\underline{0.322} & \cellcolor{gray!15}\textbf{0.385}
& \cellcolor{gray!15}\textbf{0.567} & \cellcolor{gray!15}\textbf{0.606}
& \cellcolor{gray!15}\underline{0.591} & \cellcolor{gray!15}\underline{0.622}
& \cellcolor{gray!15}\textbf{0.415} & \cellcolor{gray!15}\textbf{0.489}
& \cellcolor{gray!15}\underline{0.451} & \cellcolor{gray!15}\underline{0.513} \\

\midrule

\multirow{4}{*}{\shortstack{Unseen-Cross}}
& DDGM-Acc~\cite{deja2023diffusion}
& \underline{0.362} & \underline{0.379}
& \underline{0.422} & 0.435
& \textbf{0.416} & \textbf{0.475}
& \underline{0.284} & 0.335
& \underline{0.424} & \underline{0.470} \\

& DART
& 0.278 & 0.304
& 0.413 & 0.439
& 0.285 & 0.325
& 0.222 & 0.289
& 0.249 & 0.340 \\
& AccentBox
& 0.311 & 0.317
& 0.402 & \underline{0.456}
& 0.353 & 0.377
& 0.270 & \underline{0.357}
& 0.348 & 0.375 \\

& \cellcolor{gray!15}\textbf{Joycent (Ours)}
& \cellcolor{gray!15}\textbf{0.400} & \cellcolor{gray!15}\textbf{0.414}
& \cellcolor{gray!15}\textbf{0.445} & \cellcolor{gray!15}\textbf{0.472}
& \cellcolor{gray!15}\underline{0.410} & \cellcolor{gray!15}\underline{0.460}
& \cellcolor{gray!15}\textbf{0.377} & \cellcolor{gray!15}\textbf{0.423}
& \cellcolor{gray!15}\textbf{0.427} & \cellcolor{gray!15}\textbf{0.471} \\

\bottomrule
\end{tabular}
\end{table*}

We conduct objective evaluation under three settings, with 1,680 samples synthesized for each setting. For each target accent, the unseen-cross setting uses 24 unseen speakers from the other 6 accents (4 per accent), while the seen-inherent and unseen-inherent settings use 24 seen and unseen speakers, respectively, from the target accent. A total of 1,680 texts are randomly sampled from the AISHELL-3 test set and used across all three settings, while the accent prompt speech for each sample is randomly selected from the target accent. 

Table~\ref{tab:tts_objective} presents the overall objective results. We exclude Guangdong and Shanghai from the objective evaluation because their datasets contain dialectal variation in addition to regional accent variation, which may confound accent-specific evaluation. \textbf{Joycent shows the strongest overall accent transfer performance.} It achieves the highest WU-CS across all three settings and the highest W-CS in two settings, while remaining comparable to DDGM-Acc in the other. Notably, the consistent gains in WU-CS, which reduces the influence of speaker information in accent embeddings, provide stronger evidence that the improvement stems from accent transfer rather than speaker-related bias. \textbf{Joycent also demonstrates strong speaker preservation, }achieving the highest Spk.-CS in both unseen-speaker settings and a score comparable to the best-performing DDGM-Acc under Seen-Inherent. In terms of intelligibility, DART achieves the lowest CER across all three settings, while Joycent consistently ranks second among the synthesized systems. However, DART's lower CER is accompanied by substantially lower accent similarity, suggesting a trade-off between pronunciation clarity and target-accent realization. Taken together, \textbf{Joycent achieves the best trade-off across the three evaluation dimensions,} combining strong accent transfer and speaker preservation with competitive intelligibility. Table~\ref{tab:tts_per_accent} further breaks down the accent similarity
results by target accent. Under Unseen-Cross, Joycent achieves the highest
W-CS and WU-CS for four of the five evaluated accents, with Tianjin being
the only exception. The improvements are particularly clear for Henan and
Wuhan, where Joycent outperforms all baseline systems on both metrics.
Notably, Joycent also achieves the highest cross-accent similarity for
Singapore Mandarin despite its relatively limited training data
(2.218 hours). Under Unseen-Inherent, the results are more mixed: Joycent
performs best on Sichuan and Wuhan, while DDGM-Acc performs better on
Tianjin and Singapore Mandarin. These results suggest that the advantage
of Joycent is particularly clear in the more challenging cross-accent
setting, where the target speaker and target accent are decoupled.

\subsubsection{Perceptual Evaluation}
\begin{figure}[t]
    \centering
\includegraphics[width=0.85\linewidth]{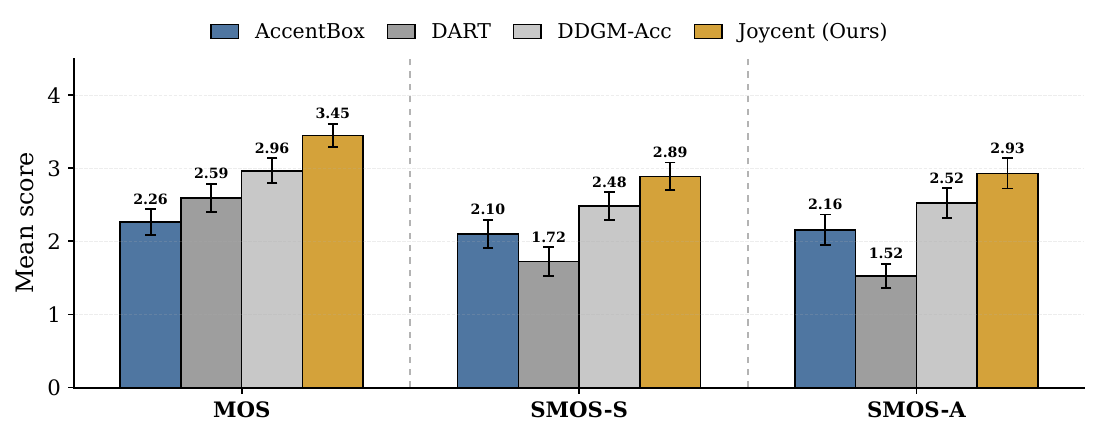}
    \caption{
    Subjective evaluation results under different models.
    We report mean opinion scores (MOS) for speech naturalness, speaker SMOS
    for speaker similarity, and accent SMOS for accent similarity.
    Error bars indicate the corresponding confidence intervals.
    }
    \label{fig:subjective_results}
\end{figure}

Fig.~\ref{fig:subjective_results} presents the overall subjective evaluation
results across the three synthesis settings.
Joycent consistently outperforms the baseline systems, achieving the highest
MOS, SMOS-S, and SMOS-A, respectively.
Compared with DDGM-Acc, the second-best system overall, Joycent improves all
three perceptual measures, indicating better speech naturalness, speaker
preservation, and accent similarity. Table~\ref{tab:listening_three_settings} further breaks down the subjective
results across the three synthesis settings.
Joycent achieves the highest MOS in all three settings, with a particularly
large advantage under Unseen-Cross, where it obtains an MOS of
$3.38{\pm}0.26$, compared with $2.62{\pm}0.26$ for the second-best system.
For speaker similarity, Joycent achieves the highest score under
Seen-Inherent and Unseen-Cross and remains close to the best-performing
system under Unseen-Inherent.
For accent similarity, Joycent achieves the highest score under both
unseen-speaker settings.
These results show that the overall perceptual advantage of Joycent remains
consistent across different synthesis conditions, particularly in the
challenging unseen-speaker cross-accent setting.

We further examine whether accent conditioning affects prosodic realization.
Fig.~\ref{fig:f0_accent_comparison} shows F0 contours of utterances
synthesized with the same text and speaker but different target accents.
F0 is extracted using WORLD~\cite{morise2016world}, and phone durations are
normalized to the maximum duration for phone-level alignment across samples.
Despite identical linguistic content and speaker identity, the generated
utterances exhibit distinct F0 trajectories across target accents,
suggesting that Joycent captures accent-dependent prosodic variation. 

\begin{table}[t]
\centering
\caption{
Subjective evaluation under three synthesis settings.
Scores are reported as mean $\pm$ 95\% confidence interval.
}
\label{tab:listening_three_settings}

\vspace{2pt}
\renewcommand{\arraystretch}{1.12}

\resizebox{\linewidth}{!}{%
\begin{tabular}{@{}llccc@{}}
\toprule
\textbf{Setting}
& \textbf{Method}
& \textbf{MOS}
& \textbf{SMOS-S}
& \textbf{SMOS-A}
\\
\midrule

\multirow{5}{*}{Seen-Inherent}
& GT
& $3.96{\pm}0.34$
& --
& --
\\
& DDGM-Acc
& \underline{$3.40{\pm}0.25$}
& \underline{$2.84{\pm}0.33$}
& $\mathbf{3.06{\pm}0.33}$
\\
& DART
& $2.76{\pm}0.41$
& $1.42{\pm}0.28$
& $1.71{\pm}0.37$
\\
& AccentBox
& $2.25{\pm}0.37$
& $1.81{\pm}0.33$
& $2.42{\pm}0.43$
\\
 
& \cellcolor{gray!15}\textbf{Joycent (Ours)}
& \cellcolor{gray!15}$\mathbf{3.54{\pm}0.29}$
& \cellcolor{gray!15}$\mathbf{3.42{\pm}0.34}$
& \cellcolor{gray!15}\underline{$2.94{\pm}0.33$}
\\

\midrule

\multirow{5}{*}{Unseen-Inherent}
& DDGM-Acc
& \underline{$2.87{\pm}0.36$}
& $\mathbf{2.69{\pm}0.36}$
& \underline{$2.53{\pm}0.42$}
\\
& DART
& $2.82{\pm}0.30$
& $1.88{\pm}0.41$
& $1.50{\pm}0.24$
\\
& AccentBox
& $2.06{\pm}0.32$
& $2.09{\pm}0.38$
& $1.84{\pm}0.29$
\\
 
& \cellcolor{gray!15}\textbf{Joycent (Ours)}
& \cellcolor{gray!15}$\mathbf{3.42{\pm}0.29}$
& \cellcolor{gray!15}\underline{$2.62{\pm}0.36$}
& \cellcolor{gray!15}$\mathbf{3.09{\pm}0.38}$
\\

\midrule

\multirow{4}{*}{Unseen-Cross}
& DDGM-Acc
& \underline{$2.62{\pm}0.26$}
& $1.91{\pm}0.30$
& $1.98{\pm}0.30$
\\
& DART
& $2.20{\pm}0.26$
& $1.86{\pm}0.33$
& $1.36{\pm}0.24$
\\
& AccentBox
& $2.48{\pm}0.20$
& \underline{$2.40{\pm}0.31$}
& \underline{$2.21{\pm}0.36$}
\\
 
& \cellcolor{gray!15}\textbf{Joycent (Ours)}
& \cellcolor{gray!15}$\mathbf{3.38{\pm}0.26}$
& \cellcolor{gray!15}$\mathbf{2.62{\pm}0.29}$
& \cellcolor{gray!15}$\mathbf{2.76{\pm}0.37}$
\\

\bottomrule
\end{tabular}%
}
\end{table}
\begin{table*}[!t]
\centering
\caption{Objective evaluation results of Joycent and ablation systems under three synthesis settings.}
\label{tab:tts_ablation}
\setlength{\tabcolsep}{3pt}
\renewcommand{\arraystretch}{1.1}

\resizebox{\textwidth}{!}{%
\begin{tabular}{lcccccccccccc}
\toprule
\multirow{2}{*}{Model}
& \multicolumn{4}{c}{Seen-Inherent}
& \multicolumn{4}{c}{Unseen-Inherent}
& \multicolumn{4}{c}{Unseen-Cross} \\
\cmidrule(lr){2-5}
\cmidrule(lr){6-9}
\cmidrule(lr){10-13}
& W-CS $\uparrow$ & WU-CS $\uparrow$ & Spk.-CS $\uparrow$ & CER $\downarrow$
& W-CS $\uparrow$ & WU-CS $\uparrow$ & Spk.-CS $\uparrow$ & CER $\downarrow$
& W-CS $\uparrow$ & WU-CS $\uparrow$ & Spk.-CS $\uparrow$ & CER $\downarrow$ \\
\midrule

\rowcolor{gray!15}
\textbf{Joycent}
& \textbf{0.490} & \textbf{0.526} & \textbf{0.626} & 0.253
& \textbf{0.469} & \textbf{0.523} & \underline{0.618} & 0.302
& \textbf{0.412} & \textbf{0.448} & 0.584 & \underline{0.258} \\

Joycent$_{\rm small}$
& 0.175 & 0.211 & 0.516 & 0.834
& 0.099 & 0.161 & 0.532 & 0.921
& 0.123 & 0.167 & 0.532 & 0.826 \\

Joycent$_{\rm large\text{-}v3\text{-}turbo}$
& 0.255 & 0.299 & 0.618 & 0.391
& 0.219 & 0.289 & \textbf{0.627} & 0.392
& 0.173 & 0.222 & \textbf{0.600} & 0.393 \\

Joycent$_{\mathrm{w/o\,GRL}}$
& 0.476 & 0.498 & 0.616 & 0.297
& 0.414 & 0.456 & 0.615 & \underline{0.288}
& 0.400 & 0.418 & \underline{0.586} & 0.267 \\

Joycent$_{\mathrm{w/o\,CLN}}$
& 0.196 & 0.219 & 0.555 & \textbf{0.154}
& 0.215 & 0.247 & 0.553 & \textbf{0.161}
& 0.208 & 0.239 & 0.545 & \textbf{0.144} \\

Joycent$_{\rm acc\text{-}blk3}$
& \underline{0.485} & \underline{0.519} & \underline{0.625} & \underline{0.246}
& \underline{0.435} & \underline{0.500} & 0.618 & 0.306
& \underline{0.404} & \underline{0.439} & 0.583 & 0.267 \\

\bottomrule
\end{tabular}%
}
\end{table*}

\subsubsection{Ablation Study}
Table~\ref{tab:tts_ablation} evaluates the contribution of the main design
choices in Joycent. Among the variants,
Joycent$_{\rm acc\text{-}blk3}$ achieves the strongest accent similarity but
remains consistently below Joycent, supporting the use of early accent
conditioning. Replacing CLN with additive conditioning
(Joycent$_{\mathrm{w/o\,CLN}}$) causes a substantial degradation in accent
similarity. Under Unseen-Cross, W-CS and WU-CS decrease from 0.412 and
0.448 to 0.208 and 0.239, respectively, demonstrating the importance of
CLN for effective accent conditioning. Removing GRL also consistently
reduces accent similarity across all three settings, showing the benefit of
speaker-disentangled accent representations for accent transfer. The encoder-scale ablation shows a similar pattern to the WhisAID
evaluation. Both Whisper Small and Whisper Large-v3-Turbo result in
substantially lower accent similarity than Whisper Medium.
Joycent$_{\rm large\text{-}v3\text{-}turbo}$ achieves the highest Spk.-CS under
Unseen-Inherent and Unseen-Cross, but this improvement is accompanied by a
large reduction in accent similarity, suggesting weaker transfer of the
target accent. Finally, Joycent$_{\mathrm{w/o\,CLN}}$ achieves the lowest
CER across all three settings despite its substantially lower accent
similarity. This suggests that lower CER alone does not necessarily indicate
better cross-accent synthesis, as insufficient target-accent transfer may
preserve pronunciations closer to the source speaker's inherent accent.






\section{Conclusion}
\label{sec:conclusion}
In this paper, we proposed Joycent, a diffusion-based accent TTS framework that uses separate accent and speaker prompts to transfer the target accent while preserving speaker identity.
Our key idea is to separate accent and speaker information in both
representation learning and TTS conditioning.
To this end, we introduced WhisAID to learn speaker-disentangled accent
representations and layer-specific conditional layer normalization to
condition the TTS model on accent and speaker information at different
stages.
We demonstrated that Joycent achieves stronger accent transfer than existing approaches while maintaining speaker similarity, with particularly clear improvements when synthesizing unseen speakers whose inherent accents differ from those provided by the accent prompts.
Subjective evaluation further shows that these improvements translate to
better perceived accent similarity and speech naturalness while preserving
speaker identity.
Our ablation studies confirm that speaker disentanglement, early accent
conditioning, and conditional layer normalization all contribute to
effective cross-accent synthesis.
These results suggest that separating accent and speaker information at both
the representation and conditioning levels provides an effective direction
for accent TTS.

\bibliographystyle{IEEEbib}
\bibliography{refs}

\end{document}